\documentclass[intlimits,twoside,a4paper]{article}
\usepackage{amsmath,amssymb}
\usepackage{graphicx}
\usepackage{wrapfig}

\usepackage[T2A]{fontenc}
\usepackage[cp1251]{inputenc}

\usepackage[eqsecnum]{cmpj}

\issue{2011}{14}{1}{14001}

\doinumber{10.5488/CMP.14.14001}

\articletype{Rapid Communication}

\title[Creation probabilities of hierarchical trees]%
{Analytical and numerical studies of creation probabilities of hierarchical trees%
}
\author[]{A.I.~Olemskoi\refaddr{label1,label2},
        S.S.~Borysov\refaddr{label2}, I.A.~Shuda\refaddr{label2}}
\addresses{
\addr{label1} Institute of Applied
Physics, Nat. Acad. Sci. of Ukraine, 58~Petropavlivs'ka~Str., 40030 Sumy, Ukraine
\addr{label2} Sumy State University, 2~Rimskii-Korsakov~Str., 40007 Sumy, Ukraine
}
\authorcopyright{A.I.~Olemskoi, S.S.~Borysov, I.A.~Shuda}
 \date{Received February 18, 2011}
\begin{document}

\maketitle

\begin{abstract}
We consider the creation conditions of diverse
hierarchical trees both analytically and numerically. A connection between the probabilities to create
hierarchical levels and the probability to associate these levels into a united
structure is studied. We argue that a consistent probabilistic picture requires
the use of deformed algebra. Our consideration is based on the study of the main
types of hierarchical trees, among which both regular and degenerate ones are
studied analytically, while the creation probabilities of Fibonacci, scale-free and arbitrary trees are determined numerically.

\pacs 02.50.-r, 89.75.-k, 89.75.Fb
\end{abstract}

\section{Introduction}\label{Sec.1}

As it is shown for diverse systems, ranging from the World Wide Web~\cite{15a}
to biological~\cite{16a} and social~\cite{20a} networks, real networks are
governed by strict organizing principles displaying the following properties:
i) most networks have a high degree of clustering; ii) many networks have been
found to be scale-free~\cite{23a} which means that the probability distribution over
node degrees, being the set of the numbers of links with neighbors, follows the
power law.

A formal basis of the theory of hierarchical structures is represented by the fact that
hierarchically constrained objects are related to an ultrametric space whose
geometrical image is the Cayley tree with nodes and branches corresponding to
elementary cells and their links~\cite{Rammal}. One of the first theoretical
pictures~\cite{Huberman} has been devoted to the diffusion
process on either uniformly or randomly multifurcating trees. The consequent study
of hierarchical structures has shown~\cite{JETP_Let2000} that their evolution is
reduced to an anomalous diffusion process in ultrametric space that arrives at a
steady-state distribution over hierarchical levels, which represents the
Tsallis power law inherent to non-extensive systems~\cite{Tsallis}. The principal
peculiarity of the Tsallis statistics is known to be governed by a deformed
algebra~\cite{Borges}.

This paper briefly represents the results of our study of creation conditions of a vast
variety of hierarchical trees on the basis of methods initially
developed within the quantum calculus~\cite{QC}. An extended version of our analysis
is published elsewhere~\cite{OBS}. The outline of the paper is as follows. In
section~\ref{Sec.2}, we discuss the statistical peculiarities of the picture of
hierarchical structure creation to demonstrate that effective energies of
hierarchical levels remain to be additive values, while the set of corresponding
probabilities becomes both non-additive and non-multiplicative due to the
coupling between different levels. Further consideration is based on an analytical and numerical study of the main types of hierarchical trees in section~\ref{Sec.4.0}. Section~\ref{Sec.8} is devoted to the discussion of obtained results.

\section{Statistical peculiarities of hierarchical ensembles}\label{Sec.2}

As pointed out above, the stationary creation probability of the $l$-th
hierarchical level takes the Tsallis form
\begin{equation}
p_l=p_0\exp_q\left(-\frac{\varepsilon_l}{\Delta}\right),\qquad \exp_q(x):=
\left[1+(1-q)x\right]_+^{1\over 1-q},\qquad [y]_+\equiv \max(0,y).
 \label{1}
\end{equation}
Here, $p_0$ is the top-level probability fixed by the normalization condition,
$q\geqslant 0$ is a deformation parameter, $\varepsilon_l$ is an effective energy of
the $l$ number, $\Delta$ is an effective temperature.
Although the energy is a key concept of the network optimization theory, it is
not always possible to match its value to a given graph. However, basing on
heuristic ideas, it is always possible to attach an effective value of energy
to some phenomenological parameter. Also, for our purpose it is convenient
to consider the nodes of the hierarchical tree as particles of a
statistical ensemble, while its edges represent couplings between these
particles.

 In contrast to the
statistical theory of complex networks~\cite{Newman}, the hierarchical systems
under our consideration cannot simultaneously display the properties of
additivity of effective energies and multiplicativity of related probabilities.
The cornerstone of our approach is that the creation of a hierarchical
structure does not break the law of the energy conservation, so that the energies
$\varepsilon_l$ remain to be additive values:
\begin{equation}
\epsilon_n:=\sum_{l=0}^{n}\varepsilon_l.
 \label{2}
\end{equation}
Within the statistical theory of random networks~\cite{Newman}, effective
energies $\varepsilon_l$ are reduced to a constant for microcanonical ensemble
and are fixed by the set of particular probabilities $p_l$ according to the relation
$\varepsilon_l=-\Delta\ln(p_l)+{\rm const}$ for both canonical and grand
canonical ensembles with an effective temperature $\Delta$.
On the other hand, due to the coupling between different levels, the hierarchy
essentially deforms the corresponding probabilities $p_l$, which become
non-multiplicative. Indeed, the probability $P_n$ to create an $n$-level
hierarchical structure is connected with total energy $\epsilon_n$ by means of
the relation $\epsilon_n=-\Delta\ln_q(P_n)$ with the deformed logarithm
$\ln_q(x):=\left(x^{1-q}-1\right)/(1-q)$. Then, the condition (\ref{2}) leads
to the additivity of these logarithms:
$\ln_q(P_n)=\sum_{l=0}^{n}\ln_q(p_l)$, and one obtains the probability
relation
\begin{equation}
P_n:=p_0\otimes_q p_1\otimes_q p_2\otimes_q\dots\otimes_q p_n,
 \label{5}
\end{equation}
where the deformed product is defined as $x\otimes_q
y:=\left[x^{1-q}+y^{1-q}-1\right]_+^{1\over 1-q}$. Thus, in contrast to
ordinary statistical systems, the creation probability $P_n$ of a hierarchical
structure is equal to the {\it deformed} production of specific probabilities
$p_l$ related to the levels $l=0,1,\dots,n$.

The above law of the deformed multiplicativity determines the probabilities
$p_l$ to create a set of hierarchical levels simultaneously. Another problem
emerges when we consider the connection between the creation probability of a
given hierarchical level $l$ and the same for each node at this level. For simplicity let us
consider a regular tree, whose nodes multifurcate to generate a
set of the $N_l$ nodes determined with inherent probabilities $\pi=p_0/N_l$
where $p_0$ is their top magnitude being a normalization constant. If one
permits additivity of the node probabilities, we arrive at the total
probability of the $l$ level realization to be independent of their numbers:
$p_l:=N_l\pi=p_0$. Since the probability $p_l$ to create a hierarchical level
decays with the level number $l$, we are forced again to replace the
trivial additive connection of the level probability $p_l$ with the node value
$\pi$ by a deformed sum.

Finally, since the creation probabilities of the hierarchical levels go beyond
probabilities related to non-hierarchical structures, the standard
normalization condition based on the use of the usual sum is broken as well.
With the growth of the difference $|q-1|$, the probability~(\ref{1}) increases at
arbitrary values of the energy $\varepsilon_l$ with respect to the non-deformed
value related to the parameter $q=1$. On the other hand, the deformed sum
$x\oplus_q y:=x+y+(1-q)xy$ decreases with the growth of the parameter $q>1$. As a
result, one can anticipate that a self-consistent probabilistic picture of
hierarchical ensembles is reached if one proposes the normalization condition
\begin{equation}
p_0\oplus_q p_1\oplus_q\dots\oplus_q p_n=1,\qquad q>1
 \label{1a}
\end{equation}
that is deformed to fix the top level probability $p_0$.

Taking into account the above statements, one obtains an explicit form of the
creation probability of a hierarchical structure~\cite{OBS}
\begin{equation}
P_n=\exp_q\left[\frac{\sum_{l=0}^{n}p_l^{1-q}-(n+1)}{1-q}\right]
=\left(\sum_{l=0}^{n}p_l^{1-q}-n\right)_+^{\frac{1}{1-q}}.
 \label{11}
\end{equation}
The relations~(\ref{11}) mean the decrease of the creation probability with the
growth of the hierarchical tree in accordance with the difference equation
\begin{equation}
P_{n-1}^{1-q}-P_n^{1-q}=1-p_n^{1-q}.
 \label{13}
\end{equation}
In the non-deformed limit $q\to 1$, relations~(\ref{5}) and~(\ref{11}) are reduced
to the ordinary rule $P_n=\prod_{l=0}^n p_l$ (respectively, equation~(\ref{13})
reads $P_{n}/P_{n-1}=p_n$), while at $q=2$ the creation probability~(\ref{11})
takes a maximal value.

According to equation~(\ref{11}) the subsequent step in the definition of the
creation probability $P_n$ of a hierarchical structure is the determination of
a set of probabilities $\{p_l\}_0^n$ related to different hierarchical levels.

\section{Level probabilities for different hierarchical trees}\label{Sec.4.0}

First, we consider a regular tree whose nodes multifurcate at the level
$l$ with constant branching index $b>1$ to generate a set of the $N_l=b^l$
nodes determined with inherent probabilities $\pi=p_0/N_l=p_0b^{-l}$.
Within naive proposition, one
could permit additivity of the node probabilities to arrive at the total
probability of the $l$ level realization to be $p_l:=N_l\pi=p_0$. Thus, within
the condition of additivity of the node probabilities, the related values
$p_l=p_0=(n+1)^{-1}$ for all levels appear to be independent of their numbers
$l=0,1,\dots,n$.

To avoid this trivial situation, we propose to replace the
common additive connection of the level probability with the node value $\pi$
by the deformed one. Such deformation leads to the required level distribution
in the binomial form~\cite{OBS}
\begin{equation}\label{aa}
p_l=\frac{[1+(1-q)b^{-l}]_+^{b^l}-1}{1-q}p_0\,.
\end{equation}
This probability increases with the growing number $l$ of hierarchical level at
$q<1$ and decays at $q>1$. From the physical point of view, the creation
probability of a lower hierarchical level should be less than for higher
levels, so that one ought to conclude that only the case $q>1$ is meaningful.
Characteristically, the form of this distribution very weakly depends on both
deformation parameter $q$ and branching index $b$ excluding the domain
$2-q\ll1$, where the probability does not decay that fast at small values of the
branching index $b$. With large growth of the parameter $b\gg1$ or level number
$l$, the dependence $p_l$ decreases faster to exponentially  reach the
minimum value
\begin{equation}
p_\infty=\frac{{\rm e}^{1-q}-1}{1-q}p_0=p_0\ln_q{\rm e}.
 \label{c}
\end{equation}

There is a distinctive feature in the behavior of the regular hierarchical tree
near the limit value $q=2$ where the dependence~(\ref{aa}) has no singularity.
This feature is corroborated with the dependence of the top level
probability $p_0$ on the
deformation parameter.
This probability increases monotonously with the $q$-growth to reach sharply
the limit value $p_0=1$ in the point $q=2$. Obviously, this means an anomalous
increase of probabilities $p_l$ for the whole set of hierarchical levels.
Though, within the domain $2-q\ll1$, the ordinary normalization condition
$\sum_{l=0}^n p_l=1$ is violated appreciably, the definition of the deformed sum
shows that the deformed normalization condition~(\ref{1a}) can be recovered with the
$q$ parameter growing. However, beyond the border $q=2$, this condition is not
satisfied at all. As a result, we arrive at the conclusion that physically
meaningful values of the deformation parameter are concentrated within the
domain $q\in[1,2]$.

The difference between regular and degenerate trees is that {\it all} nodes
multifurcate at each level in the former case, while the {\it only one} node
branches in the latter. In this sense, the degenerate tree can be considered as
an antipode of the regular one to be studied analytically. Taking into account
this peculiarity, the creation probability of the $l$th hierarchical level
takes the form~\cite{OBS}
\begin{equation}
p_l=\frac{\left[1+(1-q)b^{-l}\right]
\prod_{m=1}^l\Bigl[1+(1-q)b^{-m}\Bigl]^{b-1}-1}{1-q}p_0\,.
 \label{S6}
\end{equation}
Similarly to the case of the regular tree, this distribution decays
exponentially fast to the limit probability $p_\infty$ determined by equation~(\ref{c}).

Above, we have considered two conceptual examples of hierarchical trees with
self-similar structure, i.e., regular and degenerate trees. By contrast,
a scale-free tree has rather random structure, but the probability distribution
over hierarchical levels tends to the power-law form inherent in self-similar
statistical systems. In this case, the probability distribution over tree
levels is determined by the discrete difference equation~\cite{JETP_Let2000}
\begin{equation}
p_{l+1}-p_l=-p_l^q/\Delta,\qquad l=0,1,\dots,n
 \label{12a}
\end{equation}
accompanied with the deformed normalization condition (\ref{1a}) ($\Delta$
being a distribution dispersion).

\begin{figure}[!htb]
\centering
\hspace{5mm} (a) \hspace{65mm} (b)\\
\includegraphics[width=70mm,angle=0]{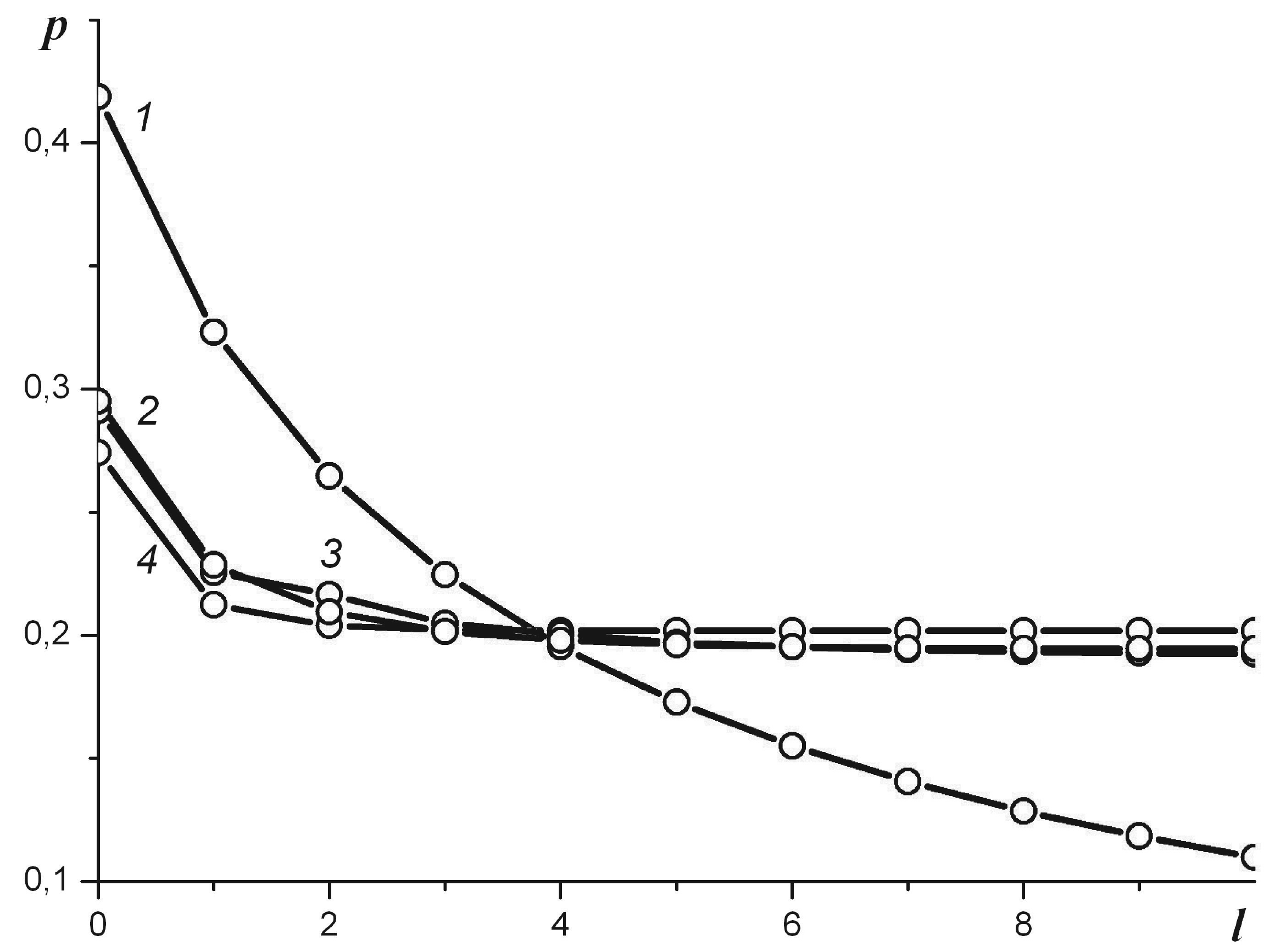}
\includegraphics[width=70mm,angle=0]{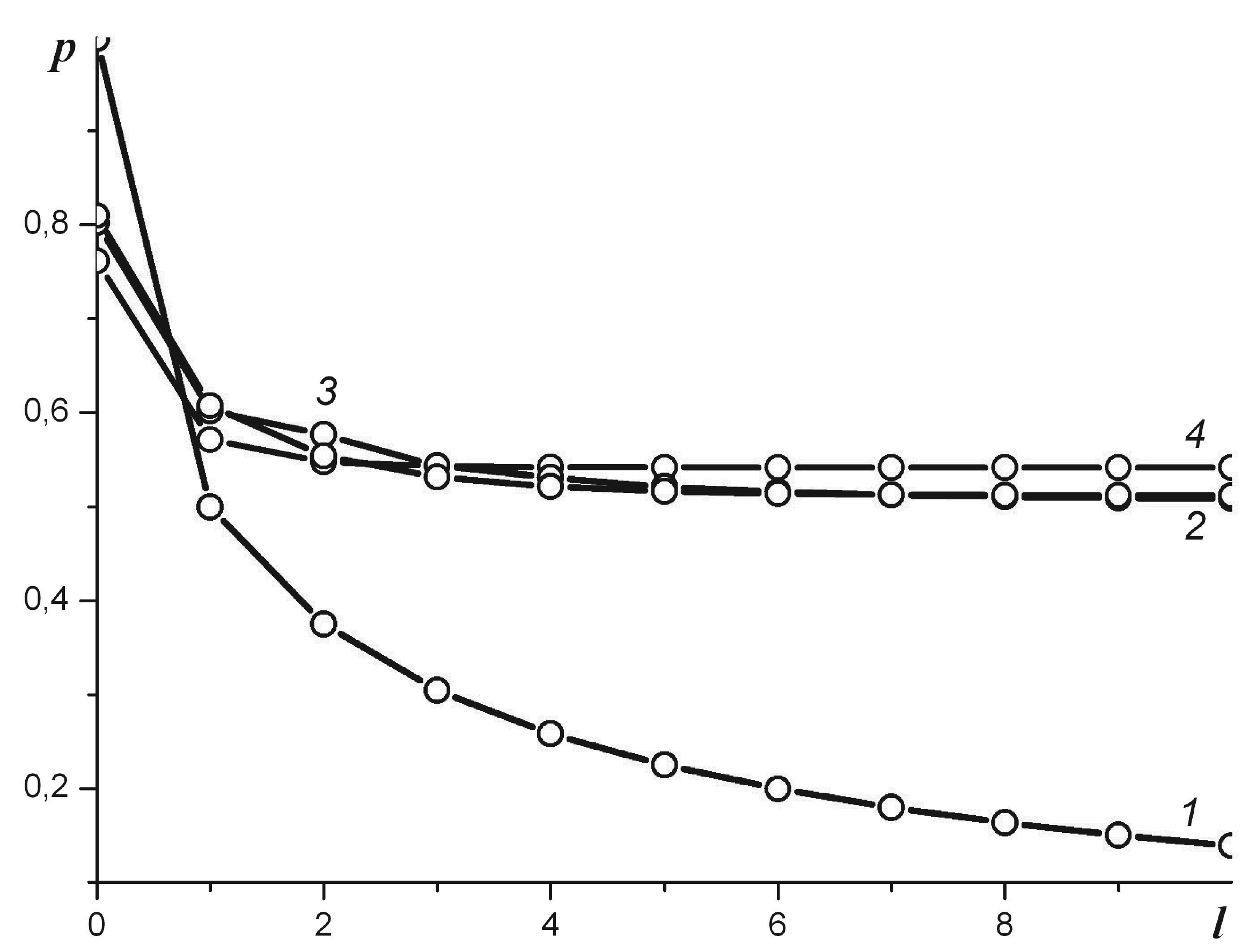}\\
 \caption{Probability distributions over hierarchical levels for scale-free,
 regular, Fibonacci and degenerate trees (curves 1-4, respectively) at
 $\Delta=2$, $b=2$, $n=10$ and $q=1.9$ (a) and $q=1.9999$ (b).}
\label{fig1}
\end{figure}

\begin{figure}[!htb]
\centering
\hspace{5mm} (a) \hspace{65mm} (b)\\
\includegraphics[width=70mm,angle=0]{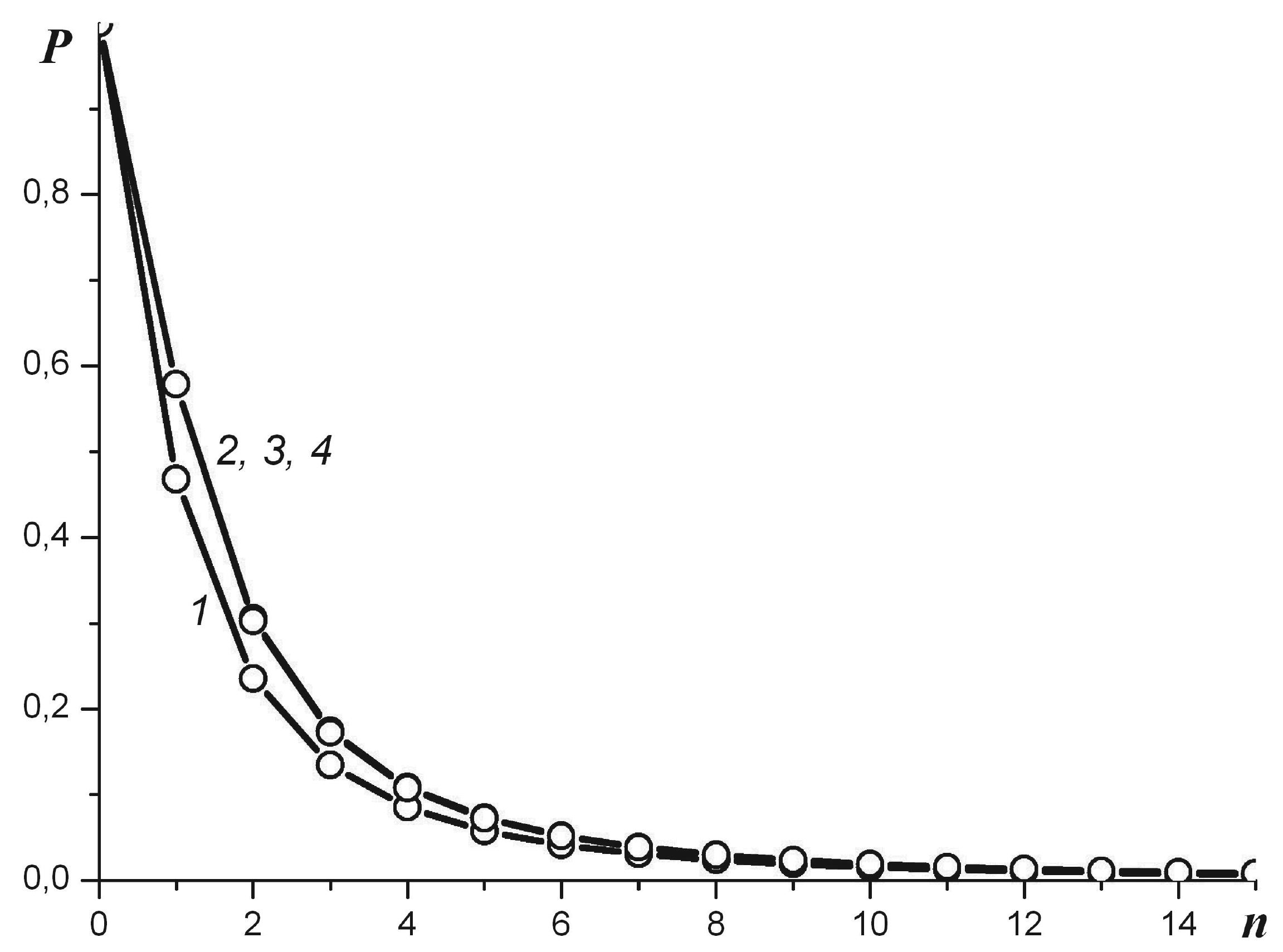}
\includegraphics[width=70mm,angle=0]{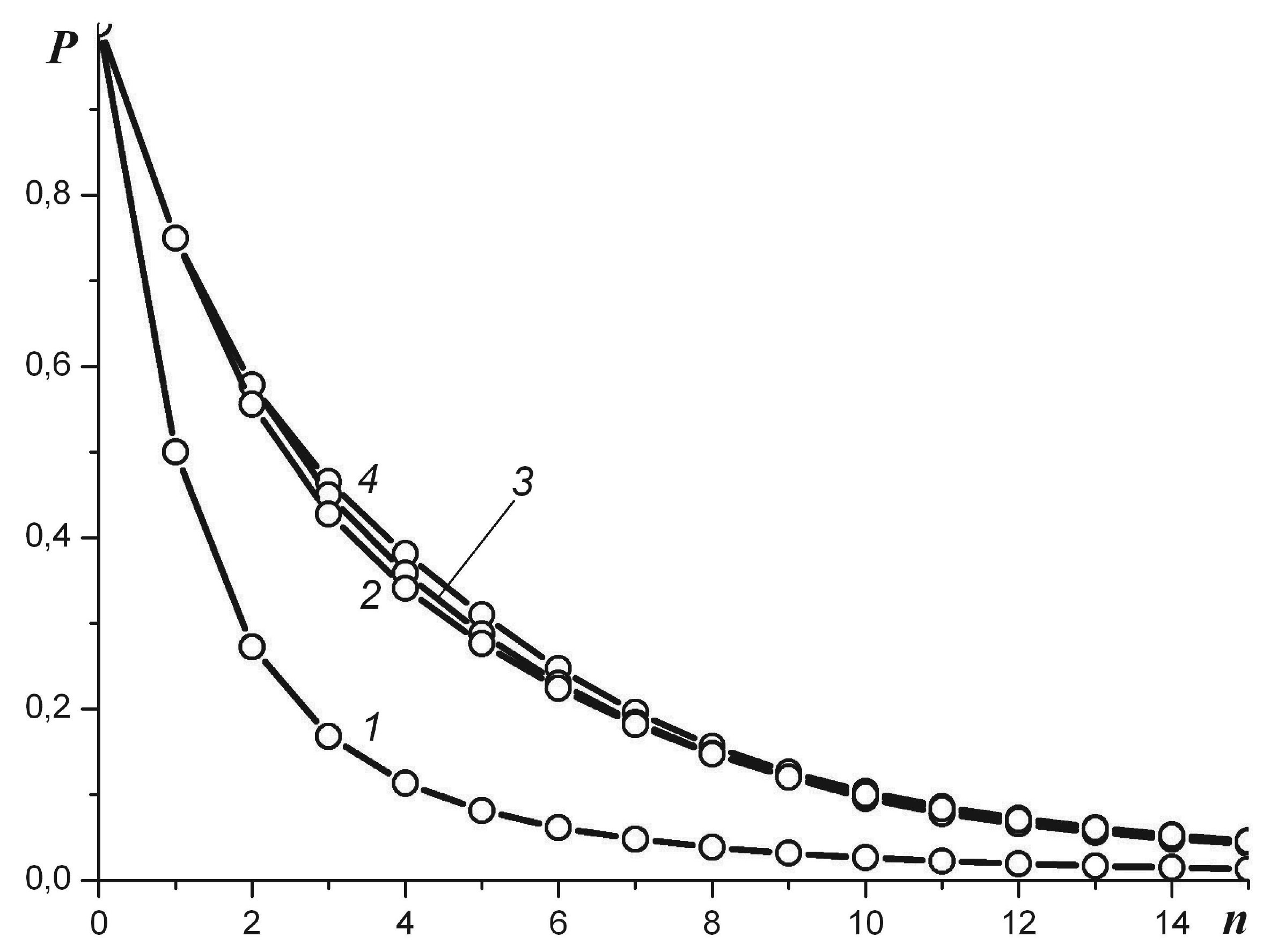}
\caption{Creation probabilities of scale-free, regular, Fibonacci
and degenerate hierarchical trees (curves 1-4, respectively) as
function of the whole level number at $\Delta=2$, $b=2$ and
$q=1.9$ (a) and $q=1.9999$ (b).} \label{fig2}
\end{figure}

In figure~\ref{fig1} we compare the probability distributions over
hierarchical levels of scale-free, regular, degenerate and Fibonacci trees at
different values of the deformation parameter. As it is seen, at all $q$-values, that the
forms of these distributions are actually similar for all the above trees excluding
the scale-free one. In the latter case, the level probability decays to zero as a power law, whereas there is a limit non-zero value~(\ref{c}) for the regular
trees. In accordance with such a behavior, the creation probabilities depicted
in figure~\ref{fig2}
decay faster for the scale-free tree than in the case of the
regular and degenerate ones. Characteristically, this difference
appears only within the domain $2-q\ll1$ of the deformation
parameter variation.

\begin{figure}%[!h]
\centering
\hspace{5mm} (a) \hspace{65mm} (b)\\
\includegraphics[width=70mm,angle=0]{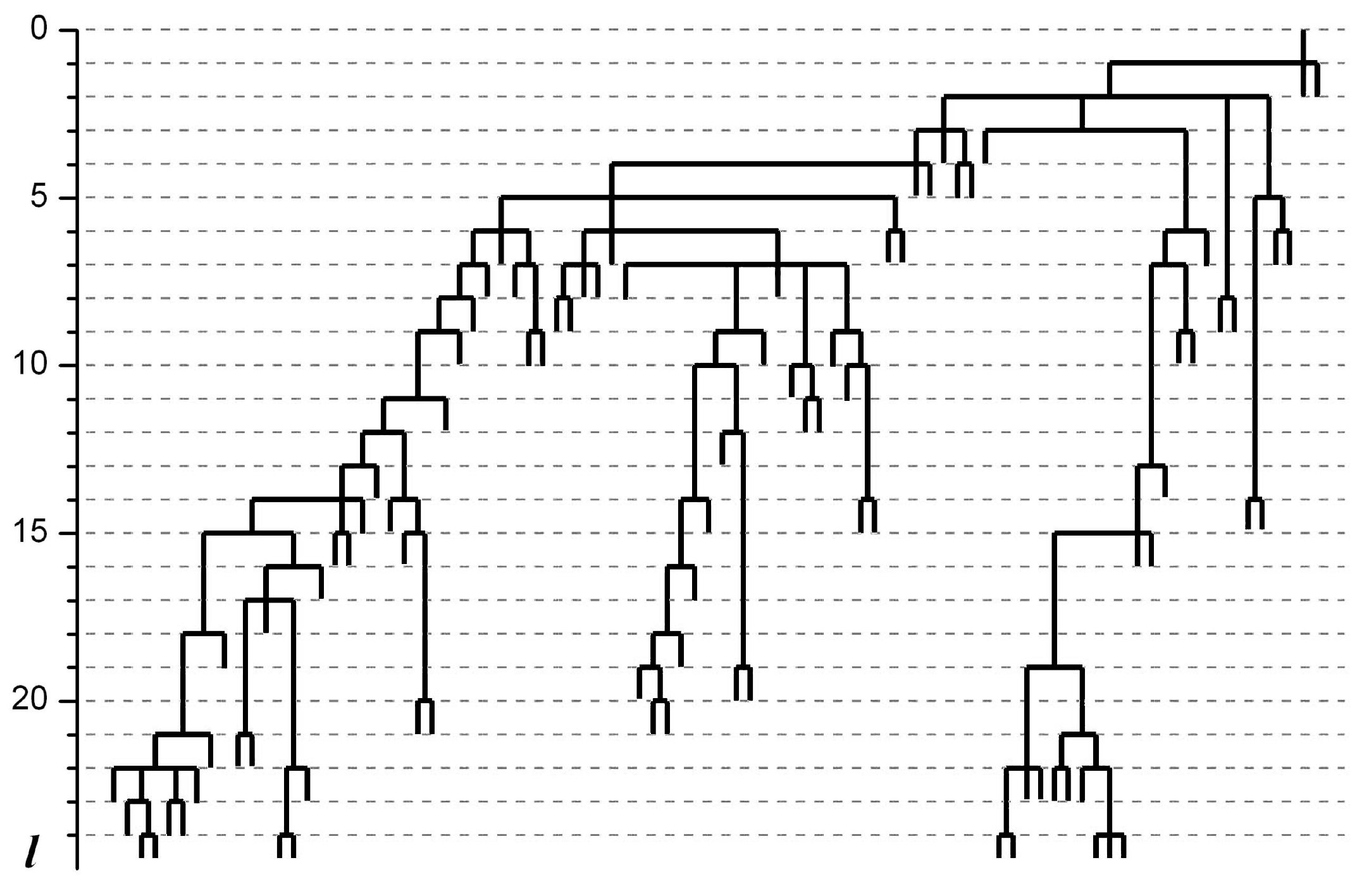} \includegraphics[width=70mm,angle=0]{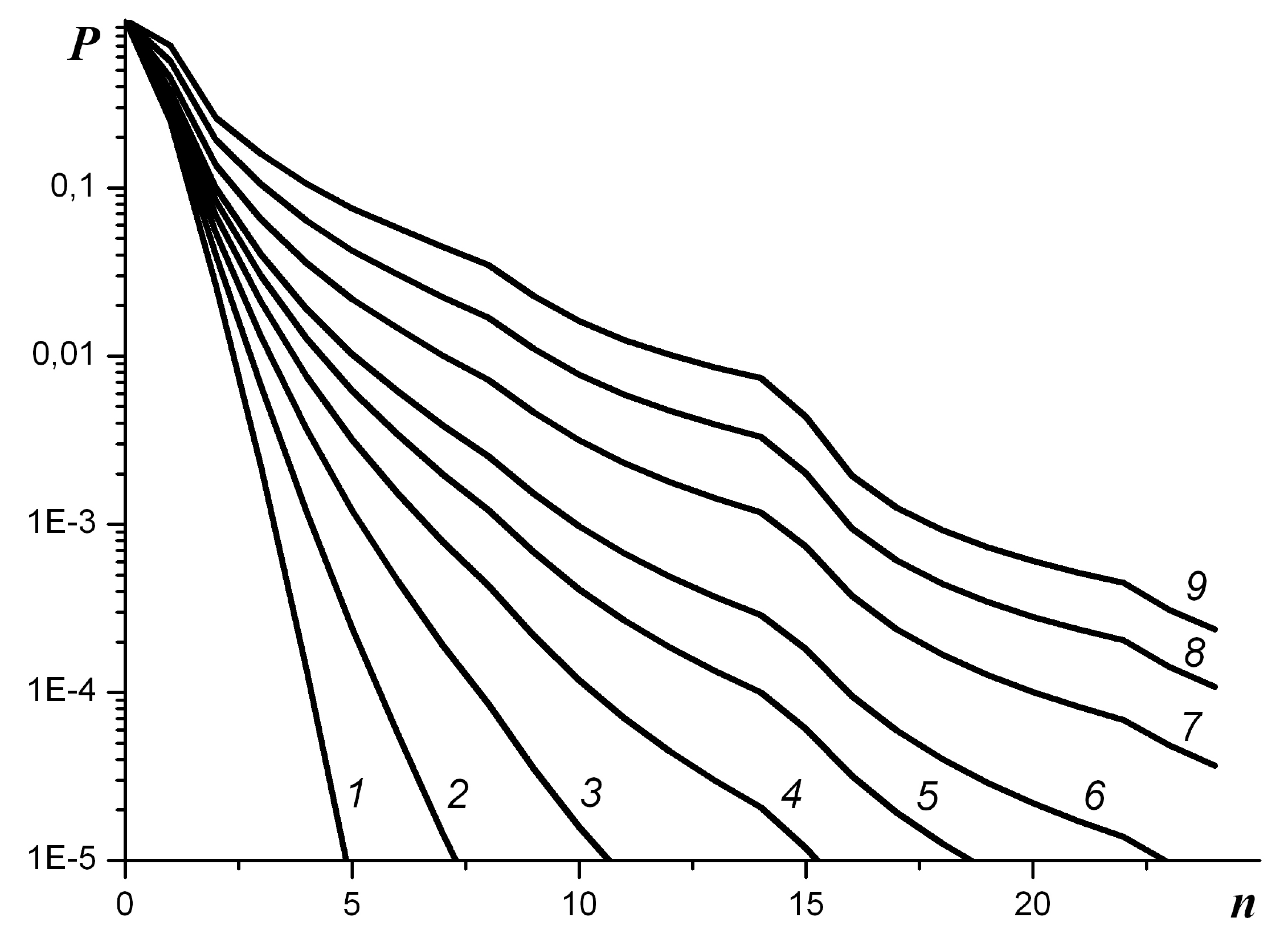}\\
  \caption{(a) Schematic representation
  of evolution tree (from Ref.~\cite{Science}). (b) Creation probability of the evolution tree vs. the level number at:
  $q = 1.0001, 1.1, 1.2, 1.3, 1.4, 1.5, 1.7, 1.9, 1.9999$ (curves
1-9, respectively).} \label{fig3.pdf}
\end{figure}

Finally, let us consider two examples of arbitrary trees,
among which the former concerns the Fibonacci tree
(number of nodes at its each level is equal to Fibonacci number), while the latter relates to the schematic
evolution tree shown in figure~\ref{fig3.pdf}a
(in the latter case, the nodes identify substantial stages in the evolution of life,
e.g., human is situated on the 24th level). Using the approach developed for
the node and level probabilities, obeying the normalization condition, we show
that the probability distributions of the Fibonacci tree depicted in figures~\ref{fig1},~\ref{fig2}
do not actually differ from the related dependencies for
both regular and degenerate trees. As concerns the evolution tree, its
probability distributions (figure~\ref{fig3.pdf}b) show that the presence of the
stopped branches (type of two rightmost ones in figure~\ref{fig3.pdf}a)
considerably decreases the creation probability of new hierarchical levels.
Particularly, the probability of human appearance takes the values greater than
$10^{-4}$ only at the deformation parameter $q=1.9999$.

\section{Concluding remarks}\label{Sec.8}

To avoid ambiguities it is worthwhile to stress that our consideration
concerns rather the probabilistic picture of creation of the hierarchical trees
themselves than hierarchical phenomena and processes evolving on these trees
(for example, hierarchically constrained statistical ensembles~\cite{13a},
diffusion processes on multifurcating trees~\cite{Huberman}, et cetera).

The principal peculiarity of the probabilistic picture elaborated is a
distinction between the deformed and non-deformed quantities. Thus, effective
energies of hierarchical levels in equation~(\ref{2}) are non-deformed quantities
because the creation of hierarchical structures does not break the conservation
law of the energy being an additive value. Moreover, the node probabilities are
determined using the non-deformed relations because these probabilities
relate to the configuration of the hierarchical tree itself (in other words,
they are determined for geometrical, rather than for probabilistic reasons). At the same
time, the hierarchy appearance essentially deforms the probability
relations~(\ref{5})--(\ref{11}) due to the coupling level probabilities $p_l$\,. Similarly, the
definition of these probabilities through corresponding node values is based on
the use of a deformed summation.

Making use of the deformed algebra leads to an increase of probabilities $p_l$ for the whole set of hierarchical levels
assuming an anomalous character near the point
$q=2$. The deformed normalization condition~(\ref{1a}) is fulfilled only at
$q\leqslant 2$, while it is broken beyond the limit $q=2$. As a result, taking into account the fact that the creation
probability of a lower hierarchical level should be less than the one for higher
levels, the physically meaningful values of the deformation parameter belong to the domain
$q\in[1,2]$.

%\newpage
\ukrainianpart

\title{Аналітичне і чисельне дослідження ймовірності утворення ієрархічних дерев}
\author{О.І. Олємской\refaddr{label1,label2}, С.С. Борисов\refaddr{label2}, І.О. Шуда\refaddr{label2}}
\addresses{
\addr{label1} Інститут прикладної фізики НАН України, вул.~Петропавлівська,~58, 40030~Суми, Україна
\addr{label2} Сумський державний університет, вул.~Римського-Корсакова,~2, 40030~Суми, Україна
}

\makeukrtitle

\begin{abstract}
\tolerance=3000%
Розглянуто аналітично і чисельно умови утворення різних
ієрархічних дерев. Досліджено зв'язок між ймовірностями утворення
ієрархічних рівнів та ймовірності об'єднання цих рівнів у єдину
структуру. Показано, що побудова послідовної ймовірнісної картини вимагає
використання деформованої алгебри. Даний розгляд заснований на дослідженні основних типів ієрархічних дерев, серед яких регулярне і вироджене досліджені аналітично, тоді як ймовірності утворення дерева Фібоначчі,
безмасштабного та довільного дерева визначені чисельно.

\keywords ймовірність, ієрархічне дерево, деформація

\end{abstract}


\begin{thebibliography}{00}

\bibitem {15a} Albert~R., Jeong H., Barab\'asi A.-L., Nature, 1999, {\bf 401}, 130;
\bibdoi{10.1038/43601}.

\bibitem {16a} Jeong H. et al., Nature, 2000, {\bf 407}, 651;
\bibdoi{10.1038/35036627}.

\bibitem {20a} Newman M.E.J., Proc. Nat. Acad. Sci. U.S.A., 2001, {\bf 98}, 404;
\bibdoi{10.1073/pnas.021544898}.

\bibitem {23a} Barab\'asi A.-L., Albert R., Science, 1999, {\bf 286}, 509;
\bibdoi{10.1126/science.286.5439.509}.

\bibitem {Rammal} Rammal R., Toulouse G., Virasoro M.A., Rev. Mod. Phys., 1986, {\bf 58}, 765;\\
\bibdoi{10.1103/RevModPhys.58.765}.

\bibitem{Huberman} Bachas C.P., Huberman B.A., Phys. Rev. Lett., 1986, {\bf 57}, 1965;
\bibdoi{10.1103/PhysRevLett.57.1965}.

\bibitem {JETP_Let2000} Olemskoi A.I., JETP Letters, 2000, {\bf 71}, 285;
\bibdoi{10.1134/1.568335}.

\bibitem{Tsallis} Tsallis C., Introduction to Nonextensive Statistical Mechanics --
Approaching a Complex World. Springer, New York, 2009;
\bibdoi{10.1007/978-0-387-85359-8}.

\bibitem {Borges} Borges E.P., Physica A, 2004, {\bf 340}, 95;
\bibdoi{10.1016/j.physa.2004.03.082}.

\bibitem{QC} Kac V., Cheung P., Quantum calculus. Springer, New York, 2002.

\bibitem{OBS} Olemskoi A., Borysov S., Shuda I., J. Phys. Stud., 2011, in press.

\bibitem{Newman} Newman M.E.J., SIAM Review, 2003, {\bf 45}, 167;
\bibdoi{10.1137/S003614450342480}.

\bibitem{Science} Sugden A.M., Jasny B.R., Culotta E., Pennisi E.,
Science, 2003, {\bf 300}, 1691;\\
\bibdoi{10.1126/science.300.5626.1691}.

\bibitem {13a} Olemskoi A.I., Ostrik V.I., Kokhan S.V., Physica A, 2009, {\bf 388}, 609;
\bibdoi{10.1016/j.physa.2008.11.019}.
\end{thebibliography}
\end{document}